\documentclass[preprint,5p,times,twocolumn]{elsarticle}

\usepackage[dvipdfm,
            CJKbookmarks=true,
            bookmarksnumbered=true,
            bookmarksopen=true,
            colorlinks=false,
            citecolor=blue,
            linkcolor=red,
            anchorcolor=green,
            urlcolor=blue
            ]{}

\usepackage{lipsum,amsmath,multicol}
\usepackage{subfigure}

\newcommand{\bea}{\begin{eqnarray}}
\newcommand{\eea}{\end{eqnarray}}
\newcommand\be{\begin{equation}}
\newcommand\ee{\end{equation}}

\begin{document}

\title{\boldmath Adiabatic Processes for Charged AdS Black Hole in the Extended Phase Space }
\author[fn1]{Shanquan Lan}
\author[fn1]{Wenbiao Liu \corref{cor1}} \ead{wbliu@bnu.edu.cn}

\address[fn1]{Department of Physics, Institute of Theoretical
Physics, Beijing Normal University, Beijing, 100875, China}

\cortext[cor1]{Corresponding author at: Department of Physics,
Institute of Theoretical Physics, Beijing Normal University,
Beijing, 100875, China}

\begin{abstract}
In the extended phase space, a general method is used to derive all the possible adiabatic processes for charged AdS black hole. Two kinds are found, one is zero temperature adiabatic process which is irreversible, the other is isochore adiabatic process which is reversible. For the zero temperature adiabatic expansion process, entropy is increasing; pressure, enthalpy, Gibbs free energy and internal energy are decreasing; system's potential energy is transformed to the work done by the system to the outer system. For the isochore adiabatic process, entropy and internal energy are fixed; temperature, enthalpy and Gibbs free energy are proportional to pressure; during the pressure increasing process, temperature is increasing and system's potential energy is transformed to its kinetic energy. Comparing these two adiabatic processes with those in normal thermodynamic system, we find that the zero temperature adiabatic process is much like the adiabatic throttling process(both are irreversible and with work done), the isochore adiabatic process is much like a combination of the reversible adiabatic process (both with fixed entropy) and the adiabatic free expansion process (both with fixed internal energy).
\end{abstract}

\maketitle
\flushbottom

\section{Introduction}
\label{intro}

Predicted by general relativity, black hole is a ``simple" object and it is an interdisciplinary research field of general relativity, quantum mechanics and thermodynamics. For these reasons, black hole has always been an interesting topic. At early times, black hole is thought to be a dead star which absorbs everything, and nothing can escape from it. While, in the 70s, Bekenstein found that a black hole possess temperature and entropy\cite{bekenstein1972,bekenstein1973}. Bardeen, Carter and Hawking established the four laws of black hole dynamics\cite{bardeencarterhawking1973}. Then Hawking found that black hole radiate and reassured the concept of black hole temperature\cite{hawkingradiation1975}. Thus the four dynamic laws become the four thermodynamic laws and black hole is believed to be a thermodynamic system. As time passes, new interesting discoveries are found, such as the Hawking-Page phase transition in the Schwarzschild-AdS black hole\cite{1983cmp} and the first order phase transition in the charged AdS black hole\cite{1999PhRvD60f4018C,1999PhRvD60j4026C}. Black hole as a thermodynamic system is further understood.

In recent years, treating the cosmological constant as a thermodynamical variable related to the dynamic pressure P ($P=-\Lambda/(8\pi)$)\cite{2012JHEP07033K}, the thermodynamic properties of various AdS black holes\cite{2013JHEP09005c,Mo2014mba,Wei2014hba,Suresh2015pra,Frassino2014pha,Caceres:2015vsa,Xu:2015rfa,Zeng:2016sei} in this extended phase space have been extensively studied. Focusing on the charged AdS black hole\cite{2012JHEP07033K,2013Spallucci1}, its thermodynamics are found to share a lot similarities with that of Van der Waals gases. For example, for both the thermodynamic systems, (1) there is an oscillating behavior on $P-V$ graph ($V$ is volume) for certain fixed $T$ which signals a phase transition phenomenon and the phase transition point can also be determined by the Maxwell's equal area law; (2) there is a swallow tail feature on $G-T$ graph ($G$ is Gibbs free energy) for certain fixed $P$ and the cross point is identified as the phase transition point. In analogy with normal fluid thermodynamic system, we have gained a much better understand of the charged AdS black hole's properties. In Ref.\cite{Wei2015iwa}, the black hole molecules are identified and are used to measure the microscopic degrees of freedom. In Ref.\cite{johnsonengine2014}, the charged AdS black hole fluid is used to design heat engines. In Ref.\cite{Okcu:2016tgt}, the Joule-Thomson expansion of charged AdS black hole fluid is studied, process of which can be used on refrigeration.

In normal thermodynamic physics, we are very interested in the fluids' adiabatic processes, as they are often used on refrigeration in our daily life. Such as the adiabatic throttling process, the reversible adiabatic process (during which the entropy $S$ is constant) and the adiabatic free expansion process (during which the internal energy $U$ is constant). In this paper, we will investigate all the possible adiabatic processes for the charged AdS black hole and compare them with those in normal thermodynamic systems, hope to gain a better understand of the charged AdS black hole's properties.

The rest of our paper is organized as follows. In section \ref{review}, we briefly introduce the first thermodynamic law and thermodynamic functions of charged AdS black hole which is established in Ref.\cite{2012JHEP07033K}. In section \ref{adiabatic}, we use a general method to find all the possible adiabatic processes for a AdS black hole thermodynamic system in canonical ensemble. For the charged AdS black hole, we find two kinds of adiabatic processes. Then their thermodynamics are studied. In the last section \ref{conclusion}, we make a conclusion and discussion.

\section{Thermodynamics of Charged AdS Black Holes in the Extended Phase Space}
\label{review}

In this section, we will give a brief review of the thermodynamics of charged AdS black holes in the extended phase space mainly based on Refs.\cite{2012JHEP07033K}. For a detail and extended knowledge of this subject, one can consult the related Refs.\cite{2013Spallucci1,2013Spallucci2,johnsonengine2014,Okcu:2016tgt,2014shaowen91,Zhao2014fea,Zhang2015ova,Wei2015iwa,Nguyen2015wfa,Lan:2015bia,Sadeghi:2016dvc,Mandal:2016anc} and references there in.

The Einstein-Maxwell bulk action reads
\begin{eqnarray}
I_{EM}=-\frac{1}{16\pi}\int d^{4}x\sqrt{-g}(R-2\Lambda-F^{2}),
\end{eqnarray}
where $\Lambda=-3/l^{2}$, the cosmological constant. A spherical charged AdS black hole is this action's solution. The metric and the $U(1)$ field are written in Schwarzschild-like coordinates as
\begin{equation}\label{metric}
ds^{2}=-f(r)dt^{2}+\frac{dr^{2}}{f(r)}+r^{2}d\Omega_{2}^{2},
\end{equation}
\begin{eqnarray}
F_{ab}=(dA)_{ab},\,\,\,\,\,\,A_{a}=-\frac{Q}{r}(dt)_{a}.
\end{eqnarray}
Here, $d\Omega_{2}^{2}$ is the element on two dimensional sphere and function $f(r)$ is given by
\begin{equation}\label{frfun}
f(r)=1-\frac{2M}{r}+\frac{Q^{2}}{r^{2}}+\frac{r^{2}}{l^{2}},
\end{equation}
with $M$ the ADM mass of the black hole, $Q$ the total charge and $l$ the AdS curvature radius.

The position of the black hole event horizon is determined as the larger root of $f(r_{+})=0$ which leads to
\begin{equation}
M=\frac{1}{2}(r_{+}+\frac{Q^{2}}{r_{+}}+\frac{r_{+}^{3}}{l^{2}}).
\end{equation}
The black hole temperature is
\begin{eqnarray}
T&=&\frac{f(r_{+})'}{4\pi}=\frac{1}{2\pi}(\frac{M}{r_{+}^{2}}+\frac{r_{+}}{l^{2}}-\frac{Q^{2}}{r_{+}^{3}})\nonumber\\
&=&\frac{1}{4\pi r_{+}}(1+\frac{3r_{+}^{2}}{l^{2}}-\frac{Q^{2}}{r_{+}^{2}}).\label{temperature}
\end{eqnarray}
The entropy is
\begin{eqnarray}
S=\frac{A}{4},\,\,\,\,\,A=4\pi r_{+}^{2},
\end{eqnarray}
and the electric potential $\Phi$, measured at infinity with respect to the horizon, is
\begin{equation}
\Phi=\frac{Q}{r_{+}}.
\end{equation}

Identifying the thermodynamic volume and the corresponding pressure as
\begin{eqnarray}
V=\frac{4}{3}\pi r_{+}^{3},\,\,\,\,\,\,P=\frac{3}{8\pi l^{2}},
\end{eqnarray}
the solution obeys the first law of black hole thermodynamics in such an extended (including P and V variables) phase space
\begin{equation}\label{firstlaw}
dM=TdS+\Phi dQ+VdP.
\end{equation}
The corresponding Smarr relation is
\begin{equation}
M=2(TS-VP)+\Phi Q,
\end{equation}
which can be derived by a dimensional scaling argument\cite{Kastor2009enth}.

The free energy of the charged AdS black hole system is the total action,
\begin{equation}
I=I_{EM}+I_{s}+I_{c}.
\end{equation}
$I_{EM}$ is the above bulk action, $I_{s}$ is a surface term\cite{2012JHEP07033K}
\begin{eqnarray}
I_{s}&=&-\frac{1}{8\pi}\int_{\partial \mathcal{M}}d^{3}x\sqrt{h}K\nonumber\\
&\,&-\frac{1}{4\pi}\int_{\partial \mathcal{M}}d^{3}x\sqrt{h}n_{a}F^{ab}A_{b},
\end{eqnarray}
and $I_{c}$ is a counter term induced to cure the infrared divergences\cite{Emparan1999con,mann1999misner}. The total action is first calculated in Refs.\cite{Chamblin1999thermo,Caldarelli2000thermo} and reads
\begin{equation}
I=\frac{\beta}{4l^{2}}(l^{2}r_{+}-r_{+}^{3}+\frac{3l^{2}Q^{2}}{r_{+}}).
\end{equation}
Associating it with the Gibbs free energy ($G$) and identifying $M$ as the enthalpy ($H$), one finds
\begin{eqnarray}
G&=&\frac{1}{4}(r_{+}-\frac{1}{l^{2}}r_{+}^{3}+\frac{3Q^{2}}{r_{+}})\nonumber\\
&=&TS-2VP+\Phi Q=M-TS=H-TS\label{gibbs1}.
\end{eqnarray}
These two thermodynamic functions are self-consistent. Then the differential formula of Gibbs free energy is easily derived from Eq.(\ref{firstlaw})
\begin{equation}\label{dergibbs}
dG=-SdT+VdP+\Phi dQ.
\end{equation}

\section{Two Kinds of Adiabatic Processes}
\label{adiabatic}

In the above section, enthalpy and Gibbs free energy have been identified. From them, the other thermodynamic functions can be easily derived. In this paper, we are interested in the internal energy
\begin{eqnarray}\label{inenergyf}
U&=&M-PV=2TS-3VP+\Phi Q\nonumber\\
&=&\frac{r_{+}}{2}+\frac{Q^{2}}{2r_{+}},
\end{eqnarray}
and its differential formula
\begin{eqnarray}
dU=TdS-PdV+\Phi dQ.
\end{eqnarray}
From all the above thermodynamic functions, we are convinced that a system with fixed $Q$ corresponds to canonical ensemble and a system with variable $Q$ corresponds to grand canonical ensemble. Since the adiabatic system is canonical ensemble, we will only consider the fixed $Q$ cases from now on,
\begin{eqnarray}\label{inenergy}
dU=TdS-PdV.
\end{eqnarray}

As we know, the change of a system's internal energy ($\Delta U$) has two sources, one is the heat absorbed by the system from the outer system ($\Delta Q_{h}$), the other is the work done by the outer system (or the minus work done by the system $-\Delta W$). Thus we have
\begin{eqnarray}\label{inenergy2}
\Delta U=\Delta Q_{h}-\Delta W.
\end{eqnarray}
Comparing with Eq.(\ref{inenergy}), we can identify
\begin{eqnarray}
dQ_{h}=TdS,\,\,\,\,\,\,\,dW=PdV.
\end{eqnarray}
During the adiabatic processes, we have
\begin{equation}
dQ_{h}=TdS=0,
\end{equation}
or equivalently
\begin{equation}
dU=-dW=-PdV.
\end{equation}

From $TdS=0$, it is easy to find that there are two kinds of adiabatic processes, one is $T=0$ process, the other is $r_{+}=const$ process.

From $dU=-PdV$, recalling the formula of $U$ in Eq.(\ref{inenergyf}), we have
\begin{eqnarray}
&\,&d(\frac{r_{+}}{2}+\frac{Q^{2}}{2r_{+}})=-4\pi P r_{+}^{2}dr_{+}\nonumber\\
\Longleftrightarrow&\,&(\frac{1}{2}-\frac{Q^{2}}{2r_{+}^{2}})dr_{+}=-4\pi P r_{+}^{2}dr_{+}.
\end{eqnarray}
One can also find that there are two cases, one is $P=\frac{1}{8\pi r_{+}^{2}}(\frac{Q^{2}}{r_{+}^{2}}-1)$, the other is $r_{+}=const$. For the first case, plug it into the temperature formula Eq.(\ref{temperature}), one get $T=0$.

So these two methods to derive all the possible adiabatic processes are equivalent. In the following section, we will analyse the derived two adiabatic processes respectively.

\subsection{Zero temperature adiabatic process}

\begin{figure}
\begin{center}
\includegraphics[width=0.38\textwidth]{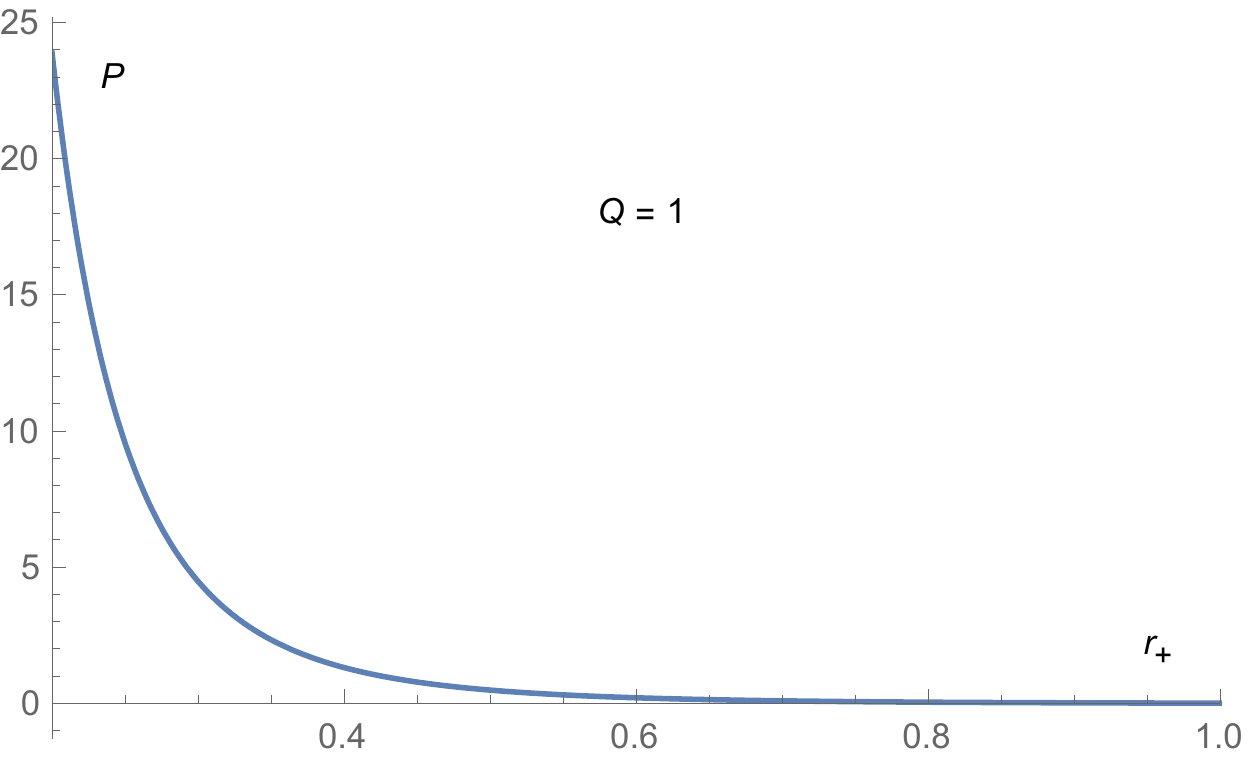}
\includegraphics[width=0.38\textwidth]{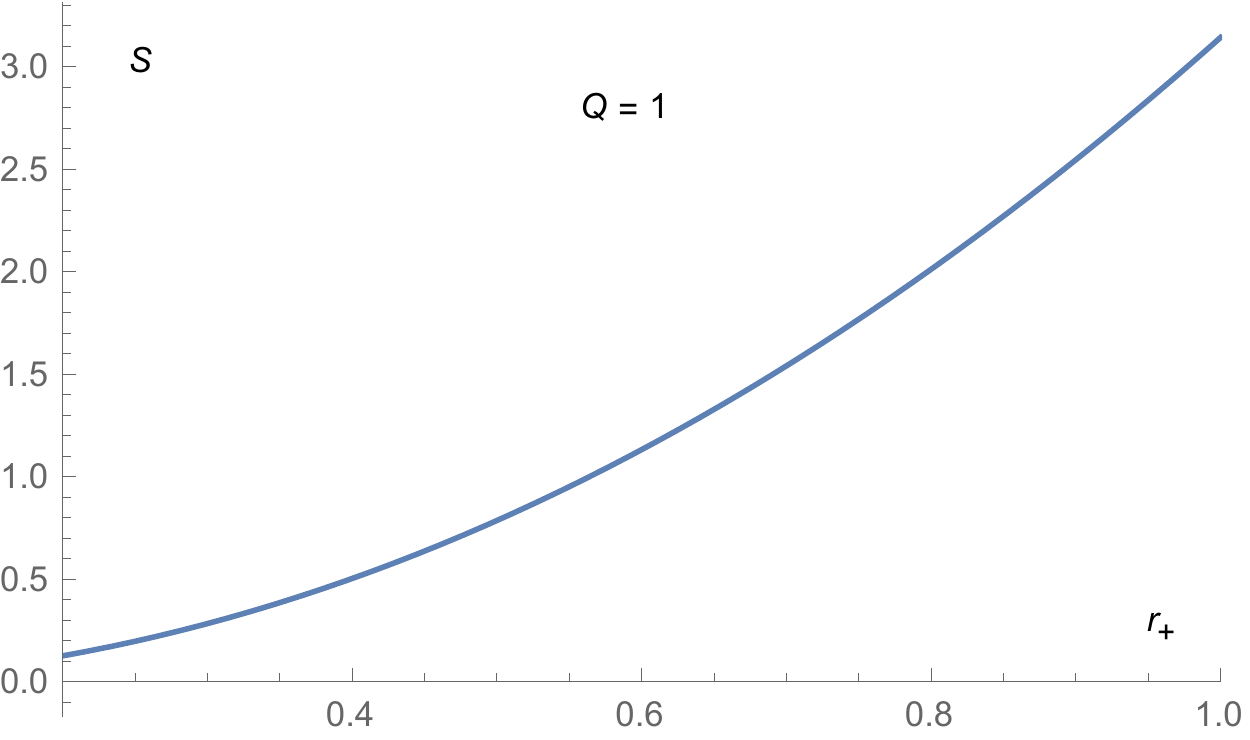}
\includegraphics[width=0.38\textwidth]{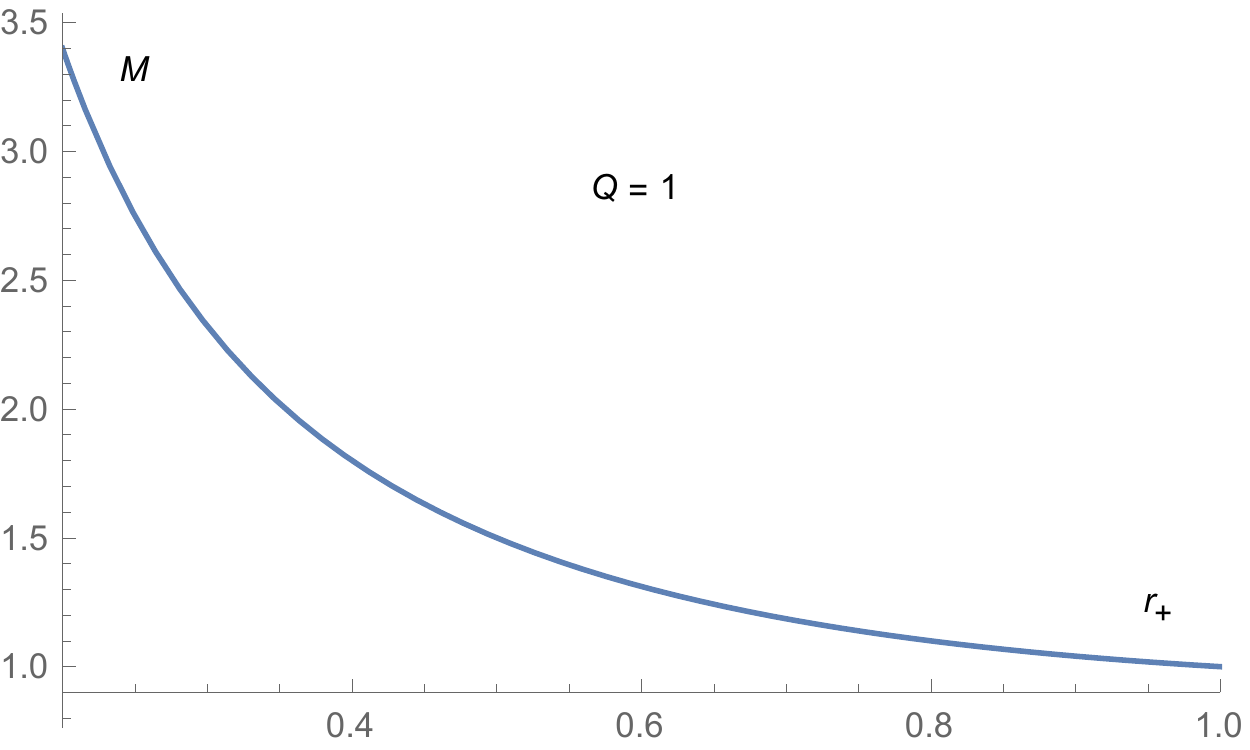}
\includegraphics[width=0.38\textwidth]{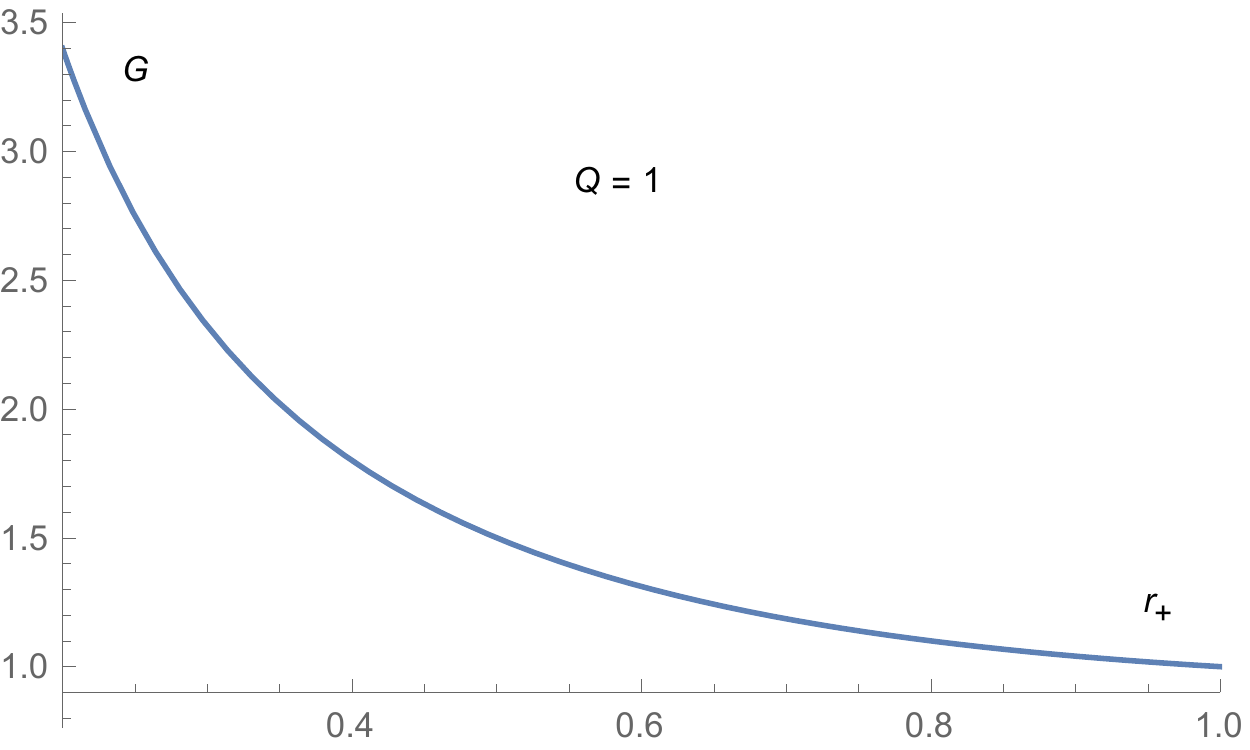}
\includegraphics[width=0.38\textwidth]{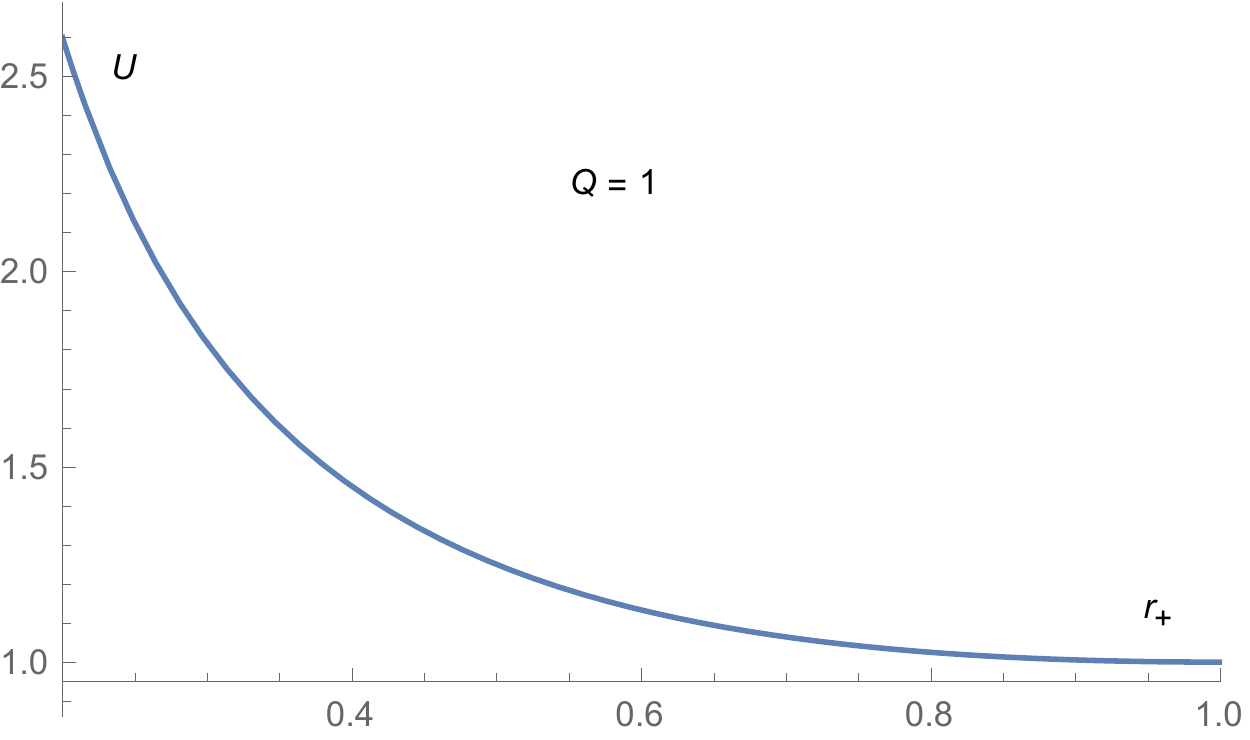}
\end{center}
\caption{The pressure $P$, entropy $S$, enthalpy $M$, Gibbs free energy $G$ and internal energy $U$ vary with respect to black hole radius $r_{+}$ for $Q=1$. For an adiabatic expansion process, we find that the entropy will increase denoting its an irreversible process; the pressure, enthalpy, Gibbs free energy and internal energy will decrease.}\label{tfun}
\end{figure}

In this case, we have already found that
\begin{eqnarray}
T=0,\,\,\,\,\,\,\,P=\frac{1}{8\pi r_{+}^{2}}(\frac{Q^{2}}{r_{+}^{2}}-1).
\end{eqnarray}
$r_{+}^{2}\leq Q^{2}$ is required to make sure that the pressure is positive. The three thermodynamic functions (enthalpy $M$, Gibbs free energy $G$ and internal energy $U$) are determined as
\begin{eqnarray}
&\,&M=\frac{r_{+}}{3}(1+\frac{2Q^{2}}{r_{+}^{2}}),\nonumber\\
&\,&G=\frac{r_{+}}{3}(1+\frac{2Q^{2}}{r_{+}^{2}}),\nonumber\\
&\,&U=\frac{r_{+}}{2}(1+\frac{Q^{2}}{r_{+}^{2}}).
\end{eqnarray}
$M$ and $G$ are found to be equal to each other. It is easy to check that during the adiabatic expansion process $0<r_{+}^{2}\leq Q^{2}$, volume $V$ and entropy $S$ are increasing functions; pressure $P$, enthalpy $M$, Gibbs free energy $G$ and internal energy $U$ are decreasing functions. We plot the pressure, entropy, enthalpy, Gibbs free energy and internal energy functions for $Q=1$ in Fig.\ref{tfun}.

For the adiabatic expansion process, the increasing of entropy denotes that it is an irreversible process; the decreasing of internal energy for fixed temperature $T=0$ denotes that the system's potential energy is transformed to the work $W$ done by the system to the outer system and it is decreasing. For the extremal state $T=P=0$, we find that $M=G=U=Q$.

Comparing with the three famous adiabatic processes (the adiabatic throttling process, the reversible adiabatic process with fixed entropy and the adiabatic free expansion process with fixed internal energy ) in normal thermodynamic system mentioned in the introduction, the adiabatic expansion process in charged black hole system is much like the adiabatic throttling process, as both processes are irreversible and the work is done by the system to the outer system during which none of entropy or internal energy are fixed. But as the temperature is already zero, this ``adiabatic throttling process" for charged black hole cannot be used on refrigeration.

\subsection{Isochore adiabatic process}

In this case, $r_{+}=const$, the quantities as functions of variable $P$ are rewritten as
\begin{eqnarray}
&\,&T=2P r_{+}+\frac{1}{4\pi r_{+}}(1-\frac{Q^{2}}{r_{+}^{2}}),\nonumber\\
&\,&S=\pi r_{+}^{2},\nonumber\\
&\,&M=\frac{r_{+}}{2}(1+\frac{Q^{2}}{r_{+}^{2}}+\frac{8\pi}{3}P r_{+}^{2}),\nonumber\\
&\,&G=\frac{r_{+}}{4}(1+\frac{3Q^{2}}{r_{+}^{2}}-\frac{8\pi}{3}P r_{+}^{2}),\nonumber\\
&\,&U=\frac{r_{+}}{2}(1+\frac{Q^{2}}{r_{+}^{2}}).
\end{eqnarray}
From the temperature formula, the positivity of $T$ and $P$ requires that $r_{+}^{2}\leq Q^{2}$ and $P\geq \frac{1}{8\pi r_{+}^{2}}(\frac{Q^{2}}{r_{+}^{2}}-1)$.

\begin{figure}
\begin{center}
\includegraphics[width=0.38\textwidth]{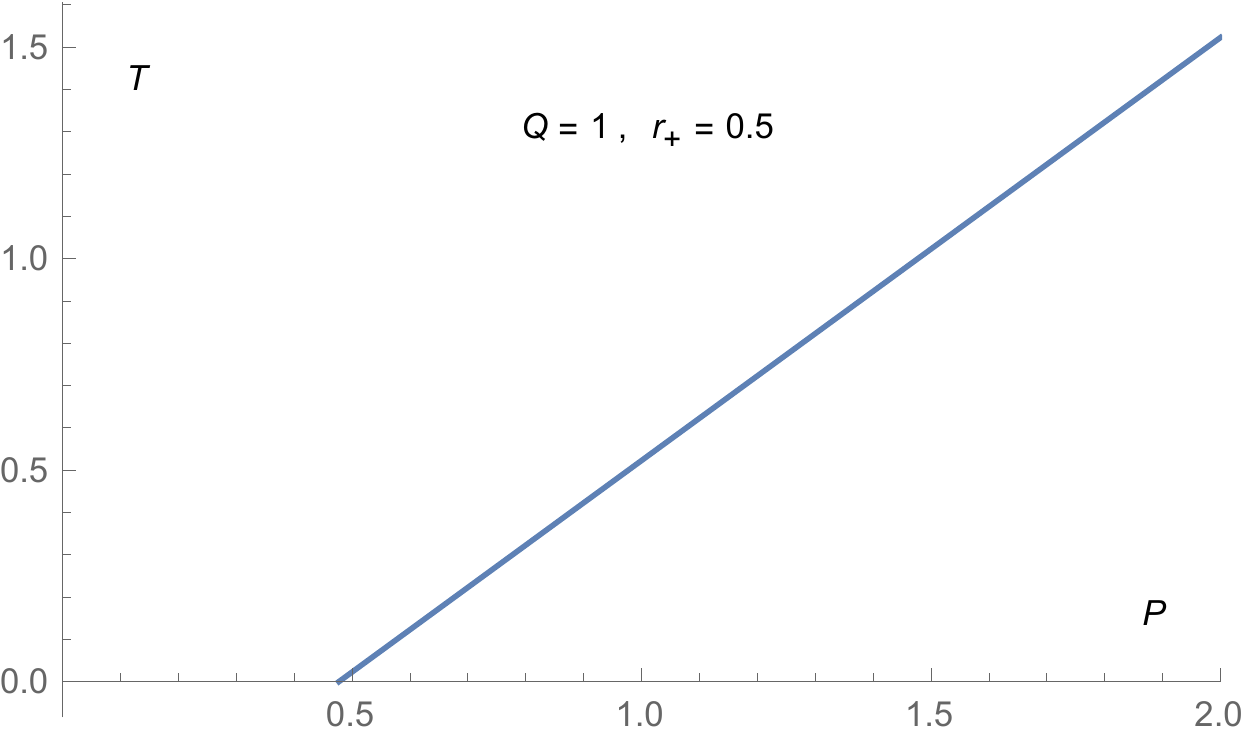}
\includegraphics[width=0.38\textwidth]{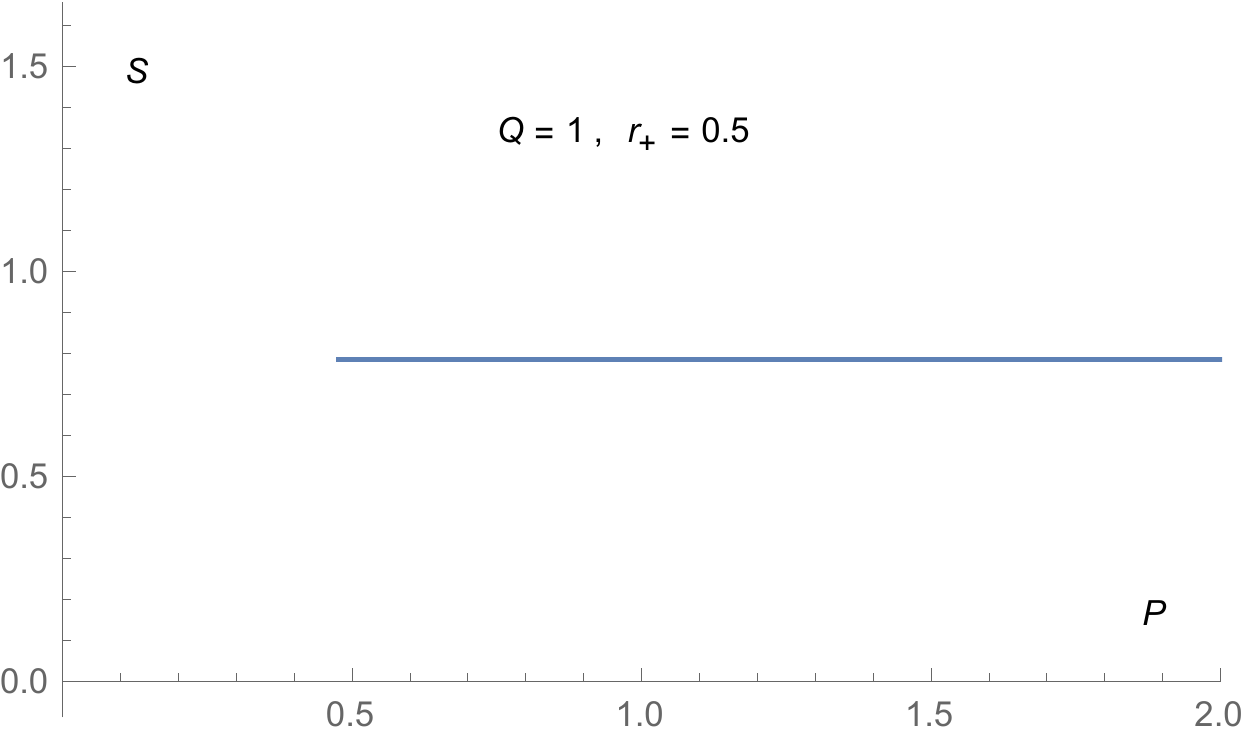}
\includegraphics[width=0.38\textwidth]{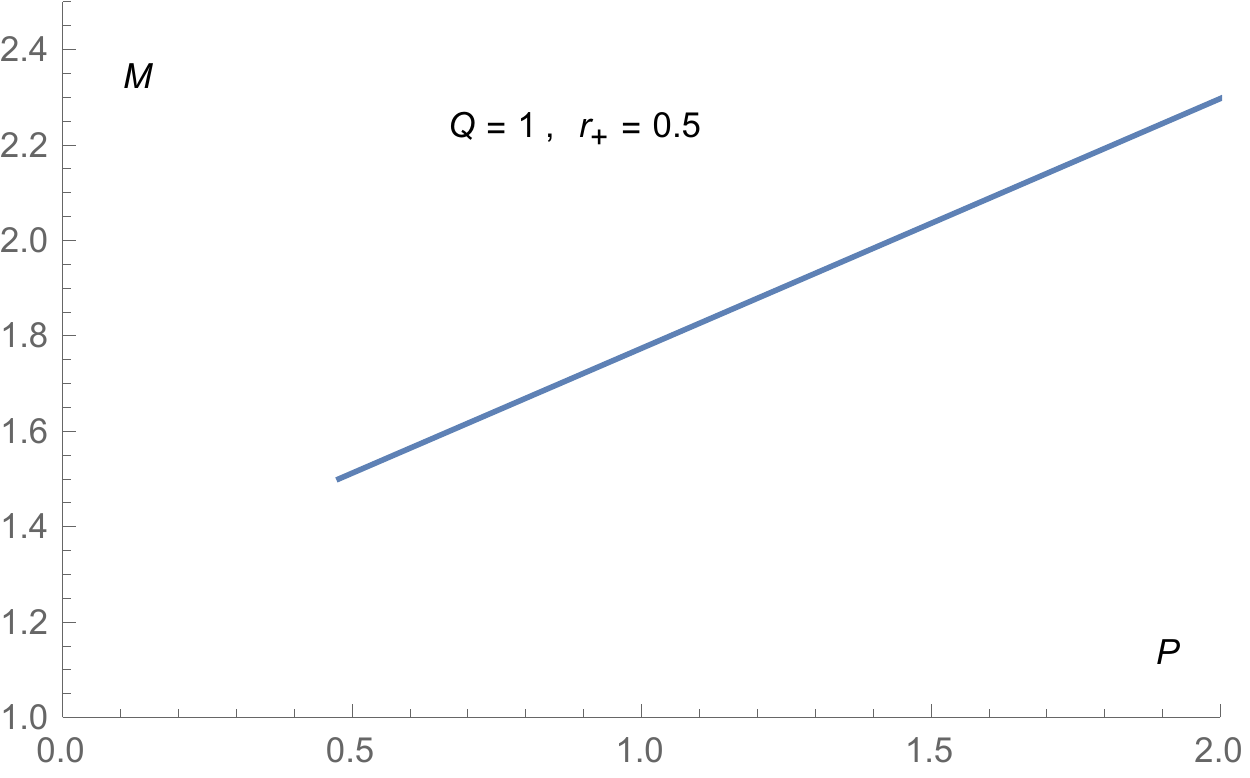}
\includegraphics[width=0.38\textwidth]{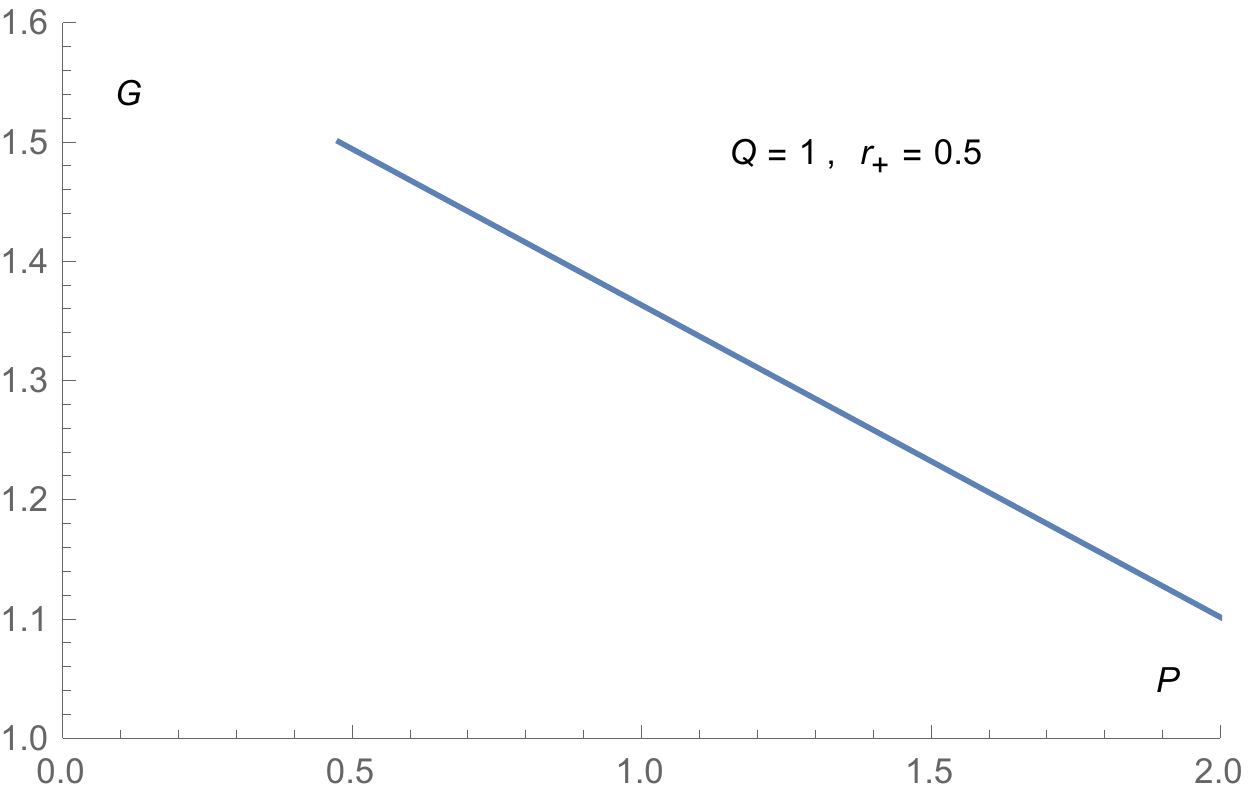}
\includegraphics[width=0.38\textwidth]{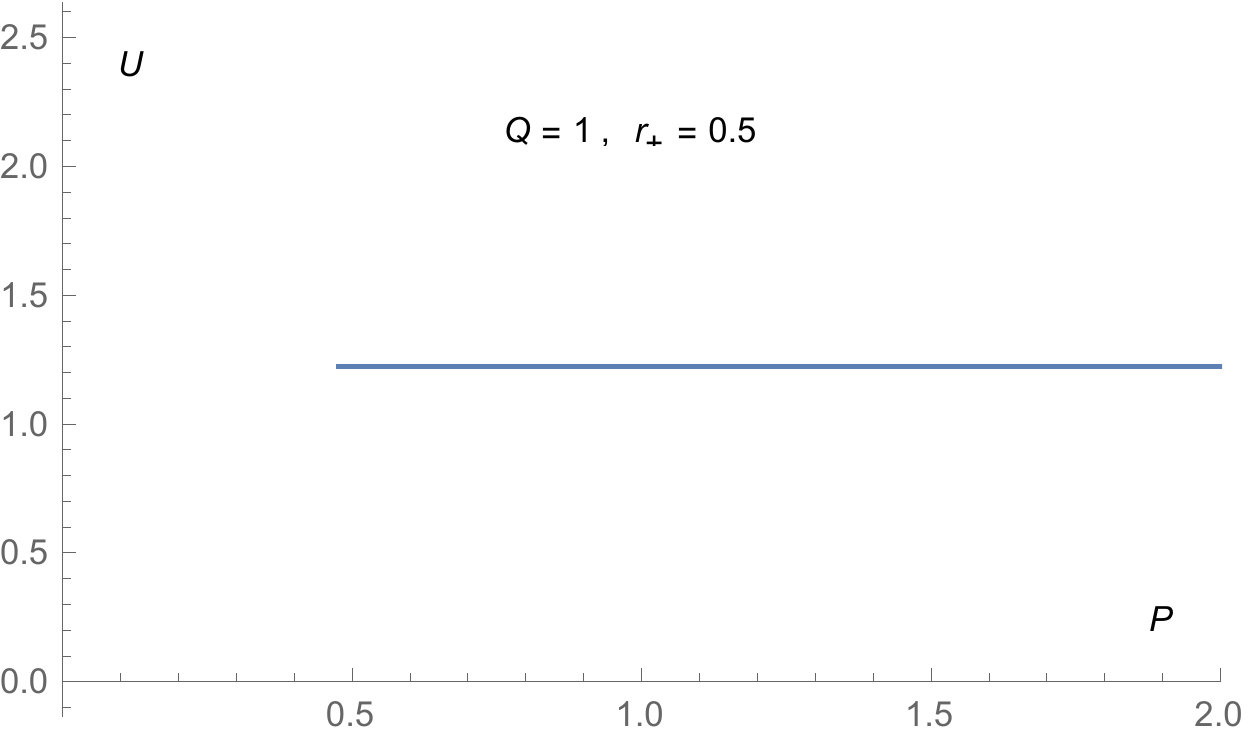}
\end{center}
\caption{The temperature, entropy, enthalpy, Gibbs free energy and internal energy vary with respect to black hole pressure $P$ for $Q=1,r_{+}=0.5$ where minimum $P_{min}\approx 0.477$. For an isochore adiabatic process, the entropy and internal energy are fixed which denotes it is a reversible process; the temperature is increasing as the increase of pressure which denotes the system's kinetic energy is increasing and as the internal energy is fixed, the potential energy is decreasing.}\label{rfun}
\end{figure}

During the isochore adiabatic process, the entropy and internal energy are fixed; the temperature, enthalpy and Gibbs free energy are all proportional to pressure. We plot the temperature, entropy, enthalpy, Gibbs free energy and internal energy functions for $Q=1,r_{+}=0.5$ which denotes $P\geq 0.477$ in Fig.\ref{rfun}.

Fixed entropy and internal energy denotes it is a reversible process. When the pressure is increasing, the temperature will increase which denotes the system's kinetic energy is increasing and as the internal energy is fixed, the potential energy is decreasing. For this process, the system's potential energy is transformed to its kinetic energy. When the pressure is decreasing, the process is reversal to the pressure increasing case and the system's kinetic energy is transformed to its potential energy.

Also comparing with the three adiabatic processes (the adiabatic throttling process, the reversible adiabatic process with fixed entropy and the adiabatic free expansion process with fixed internal energy ) in normal thermodynamic system mentioned in the introduction, the isochore adiabatic process in charged black hole system is much like a combination of the reversible adiabatic process and the adiabatic free expansion process, as the entropy and internal energy are fixed during the process. One can also say that the corresponding ``reversible adiabatic process" and ``adiabatic free expansion process" in charged black hole system are identical to each other which becomes the isochore adiabatic process. As during the isochore adiabatic process, when the pressure is decreasing, the temperature will decrease, so it can always be used on refrigeration.

\section{Conclusion and Discussion}
\label{conclusion}

Charged AdS black hole is a thermodynamic system. Treating its pressure as $P=-\Lambda/(8\pi)$, we reviewed how the first thermodynamic law and the thermodynamic functions in this extended phase space were determined in section \ref{review}. In section \ref{adiabatic}, for the canonical ensemble, focusing on the internal energy, we introduced a general method to find all the possible adiabatic processes. There are only two kinds of adiabatic processes for the charged AdS black hole, one is the zero temperature adiabatic process, the other is the isochore adiabatic process.

The zero temperature adiabatic process is irreversible. There are contraction and expansion processes which are reversal to each other. So we only investigate the expansion process. During this process, there is work done by the system to the outer system. So the system's internal energy is decreasing. As the temperature is fixed which means the system's kinetic energy is fixed, so the system's potential energy is decreasing which is transformed to the work. This adiabatic process is found to much like the adiabatic throttling process in normal thermodynamic system, except that it cannot be used on refrigeration.

The isochore adiabatic process is reversible. There are pressure increase and decrease processes. During the process, the system's entropy and internal energy are fixed. Increasing the pressure, the temperature will increase, so does the system's kinetic energy. As a result, system's potential energy is decreasing which is transformed to system's kinetic energy. Decreasing the pressure, the process is reversal to that of increasing the pressure. This adiabatic process is found to much like a combination of the reversible adiabatic process (with fixed entropy) and the adiabatic free expansion process (with fixed internal energy) in normal thermodynamic system, as the entropy and internal energy are fixed during the process. The isochore adiabatic process can always be used on refrigeration, and it is used to design heat engine in Ref.\cite{johnsonengine2014}.

In the end, we will like to emphasize on the method used to find all the adiabatic processes. The change of a canonical ensemble system's internal energy composes only two parts, one is the heat absorbed by the system from the outer system, the other is the work done by the outer system. So there are two ways to find all the adiabatic processes, one way is to consider the heat differential terms in the internal energy differential formula which should be zero, the other way is to calculate the minus work done by the system which should be equal to the change of system's internal energy. This method can be used on other kinds of AdS black hole systems. For the more complicated AdS black hole systems with more parameters, the adiabatic processes should be more abundant than just two kinds.

\hspace{2em}

\section*{Acknowledgments}

 This work is supported by NSFC with Grant Nos. 11235003, 11175019 and 11178007.

\bibliographystyle{elsarticle-num-names}
\bibliography{thereference}

\end{document}